\documentclass[12pt] {article}
\usepackage{psfrag}
\usepackage{graphicx}
\usepackage{latexsym,amsfonts}
\usepackage{amssymb}
\begin{document}
\setcounter{page}{1}
\def\theequation{\arabic{section}.\arabic{equation}}
\def\theequation{\thesection.\arabic{equation}}
\setcounter{section}{0}

\title{On the estimate of the $\sigma^{(I = 1)}_{KN}(0)$--term value
  \\ from the energy level shift of kaonic hydrogen\\ in the ground state}

\author{A. N. Ivanov\,\thanks{E--mail: ivanov@kph.tuwien.ac.at, Tel.:
+43--1--58801--14261, Fax: +43--1--58801--14299}~\thanks{Permanent
Address: State Polytechnic University, Department of Nuclear Physics,
195251 St. Petersburg, Russian Federation}\,,
M. Faber\,\thanks{E--mail: faber@kph.tuwien.ac.at, Tel.:
+43--1--58801--14261, Fax: +43--1--58801--14299}\,,
V. A. Ivanova\,\thanks{E--mail: viola@kph.tuwien.ac.at, State
Polytechnic University, Department of Nuclear Physics, 195251
St. Petersburg, Russian Federation}\,, J. Marton\,\thanks{E--mail:
johann.marton@oeaw.ac.at}\,, N. I. Troitskaya\,\thanks{State
Polytechnic University, Department of Nuclear Physics, 195251
St. Petersburg, Russian Federation}}

\date{\today}

\maketitle

\begin{center} {\it Atominstitut der \"Osterreichischen
    Universit\"aten, Technische Universit\"at Wien,\\ Wiedner
    Hauptstrasse 8-10, A-1040 Wien, \"Osterreich \\ and \\ Stefan
    Meyer Institut f\"ur subatomare Physik, \"Osterreichische Akademie
    der Wissenschaften, Boltzmanngasse 3, A-1090, Wien, \"Osterreich}
\end{center}

\begin{center}

\begin{abstract}
Using the experimental data on the energy level shift of kaonic
  hydrogen in the ground state (the DEAR Collaboration, Phys. Rev.
  Lett. {\bf 94}, 212302 (2005)) and the theoretical value of the
  energy level shift, calculated within the phenomenological quantum
  field theoretic approach to the description of strong low--energy
  $\bar{K}N$ and $\bar{K}NN$ interactions developed at Stefan Meyer
  Institut f\"ur subatomare Physik in Vienna, we estimate the value of
  the $\sigma^{(I = 1)}_{KN}(0)$--term of low--energy $\bar{K}N$
  scattering. We get $\sigma^{(I = 1)}_{KN}(0) = (433 \pm 85)\,{\rm
    MeV}$. This testifies the absence of strange quarks in the proton
  structure.  
\end{abstract}

PACS: 11.30.Rd, 13.75.Jz, 25.80.Nv, 36.10.Gv

\end{center}

\newpage

\section{Introduction}
\setcounter{equation}{0}

Recently in Refs.\cite{IV3}--\cite{IV8} (see also \cite{VI05}) we have
proposed a phenomenological quantum field theoretic model for strong
low--energy $\bar{K}N$ and $\bar{K}NN$ interactions at threshold for
the analysis of the experimental data by the DEAR Collaboration
Refs.\cite{DEAR,SIDDHARTA} on the energy level displacement of the
ground state of kaonic hydrogen
\begin{eqnarray}\label{label1.1}
  - \epsilon^{(\exp)}_{1s} + i\,\frac{\Gamma^{(\exp)}_{1s}}{2} =
  (- 193 \pm 37)
  + i(125 \pm 56)\,{\rm eV}.
\end{eqnarray} 
According to the DGBTT formula \cite{SD54}, the energy level
displacement of the ground state of kaonic hydrogen is related to the
S--wave amplitude $f^{K^-p}_0(0)$ of elastic $K^-p$ scattering at
threshold as
\begin{eqnarray}\label{label1.2}
  - \,\epsilon_{1s} + i\,\frac{\Gamma_{1s}}{2} = 
  2\,\alpha^3\mu^2 
  \,f^{K^-p}_0(0) =  412.13\,f^{K^-p}_0(0),
\end{eqnarray}
where $2\alpha^3\mu^2 = 412.13\,{\rm eV}/{\rm fm}$, $\mu = m_K
m_p/(m_K + m_p) = 323.48\,{\rm MeV}$ is the reduced mass of the $K^-p$
pair, calculated for $m_K = 493.68\,{\rm MeV}$ and $m_p = 938.27\,{\rm
  MeV}$, and $\alpha = 1/137.036$ is the fine--structure constant
Ref.\cite{PDG04}.  An accuracy of the DGBTT formula
Eq.(\ref{label1.2}) is about $3\,\%$ including the vacuum polarization
correction Ref.\cite{TE04}.

The real part ${\cal R}e f^{K^-p}_0(0)$ of the S--wave amplitude
$f^{K^-p}_0(0)$ defines the energy level shift of kaonic hydrogen in the
ground state.  It has been calculated in \cite{IV3}--\cite{VI05} up to
next--to--leading order in chiral expansion
\begin{eqnarray}\label{label1.3}
  {\cal R}e f^{K^-p}_0(0) = a^{K^-p}_0 - \frac{1}{4\pi
    F^2_{\pi} }\,\frac{\mu}{m_K}\Big[\sigma^{(I = 1)}_{KN}(0) - \frac{ m^2_K}{ 
    4 m_N} i\!\int\! d^4x \langle p
  |{\rm T}(J^{4+i5}_{50}(x)J^{4-i5}_{50}(0))|
  p\rangle\Big]\!,
\end{eqnarray}
where $a^{K^-p}_0 = (-\,0.54\pm 0.05)\,{\rm fm}$ is the S--wave
scattering length of elastic $K^-p$ scattering calculated to leading
order in chiral expansion \cite{IV8,VI05}( see also \cite{IV8a}),
including the contribution of strange baryon resonances such as the
$\Lambda(1405)$ resonance \cite{IV8} and the corrections, caused by
inelastic channels $K^- p \to \Sigma\pi$ and $K^-p\to \Lambda^0\pi^0$
\cite{VI05}, 

The two last terms in Eq.(\ref{label1.3}) are next--to--leading order
corrections in chiral expansion, where $F_{\pi} = 92.4\,{\rm MeV}$ is
the PCAC constant \cite{PDG04}. The second one is the $\sigma^{(I =
  1)}_{KN}(0)$--term of $\bar{K}N$ scattering. It is defined by
\cite{ER72}--\cite{JG99}
\begin{eqnarray}\label{label1.4}
  \sigma^{(I = 1)}_{KN}(0) = \frac{m_u + m_s}{4m_N}\,
\langle p|\bar{u}(0)u(0) + \bar{s}(0)s(0)|p\rangle,
\end{eqnarray}
where $u(0)$ and $s(0)$ are current quark fields, $m_u$ and $m_s$ are
their masses. Since the $\sigma^{(I = 1)}_{KN}(0)$--term is
proportional to $(m_u + m_s)$, so at leading order in chiral expansion
the matrix element $\langle p|\bar{u}(0)u(0) +
\bar{s}(0)s(0)|p\rangle$ should be calculated in the chiral limit.

The chiral order of the third term is defined by the squared
$K^-$--meson mass $m^2_K = O(m_u + m_s)$. This implies that the matrix
element $\langle p |{\rm T}(J^{4+i5}_{50}(x)J^{4-i5}_{50}(0))|
p\rangle$, where $J^{4\pm i5}_{50}(x)$ are time--components of axial
hadronic currents $J^{4\pm i5}_{5\mu}(x)$ changing strangeness
$|\Delta S| = 1$, as well as $\langle p|\bar{u}(0)u(0) +
\bar{s}(0)s(0)|p\rangle$ has to be calculated in the chiral limit.
  
The real part of the S--wave amplitude, given by Eq.(\ref{label1.3}),
determines the energy level shift $\epsilon^{(\rm th)}_{1s}$ of the
ground state of kaonic hydrogen. It is equal to
\cite{IV5}--\cite{IV8}
\begin{eqnarray}\label{label1.5}
  \hspace{-0.3in}\epsilon^{(\rm th)}_{1s} = \epsilon^{(0)}_{1s} + 
  \frac{\alpha^3 \mu^3}{2\pi m_K 
    F^2_{\pi}}\Big[\sigma^{(I = 1)}_{KN}(0) - \frac{ m^2_K}{ 
    4 m_N} i\!\int\! d^4x \langle p
  |{\rm T}(J^{4+i5}_{50}(x)J^{4-i5}_{50}(0))|
  p\rangle\Big],
\end{eqnarray}
where $\epsilon^{(0)}_{1s} = (238\pm 21)\,{\rm eV}$ is caused by
$a^{K^-p}_0 = (-\,0.54 \pm 0.05)\,{\rm fm}$ and the isospin--breaking
corrections, which contribute only to the energy level displacement of
the ground state of kaonic hydrogen but not to the S--wave amplitude
$f^{K^-p}_0(0)$ of elastic $K^-p$ scattering at threshold
\cite{IV7,IV8}.

The theoretical estimates of $\sigma^{(I = 1)}_{KN}(0)$, carried out
within ChPT \cite{JG83} with a dimensional regularization of divergent
integrals, converge to the number $\sigma^{(I = 1)}_{KN}(0) = (200 \pm
50)\,{\rm MeV}$ \cite{VB93}. The values of the $\sigma^{(I =
  1)}_{KN}(0)$--term, calculated in ChPT with a cut--off
regularization, are by a factor 2 larger than $\sigma^{(I =
  1)}_{KN}(0) = (200 \pm 50)\,{\rm MeV}$ \cite{VB93}.

The aim of this paper is to estimate the value of the $\sigma^{(I =
  1)}_{KN}(0)$--term from the experimental data by the DEAR
Collaboration Eq.(\ref{label1.1}) using the theoretical value of the
energy level shift given by Eq.(\ref{label1.5}).

\section{Estimate of the $\sigma^{(I = 1)}_{KN}(0)$--term}
\setcounter{equation}{0}

In order to extract the value of the $\sigma^{(I = 1)}_{KN}(0)$--term
from the experimental data on the energy level shift of kaonic
hydrogen in the ground state Eq.(\ref{label1.1}) by using the
theoretical value of the energy level shift Eq.(\ref{label1.5}) we
have to calculate the contribution of the matrix element $\langle
p\vert J^{4+i5}_{50}(x)J^{4-i5}_{50}(0)\vert p\rangle$. As has been
mentioned above, the contribution of this matrix element should be
taken in the chiral limit, i. e. to leading order in ChPT \cite{JG83}.

Following \cite{IV3,TE05} we can transcribe the third term in
Eq.(\ref{label1.5}) as follows
\begin{eqnarray}\label{label2.1}
 \hspace{-0.1in}i\!\!\int\!\! d^4x \langle p
  |{\rm T}(J^{4+i5}_{50}(x)J^{4-i5}_{50}(0))|
  p\rangle =  \frac{1}{2}\sum_{\sigma_p = \pm 1/2}\sum_X(2\pi)^3\delta^{(3)}(\vec{p}_X)
\frac{\vert \langle X\vert J^{4-i5}_{50}(0)|
  p(\vec{0},\sigma_p)\rangle\vert^2}{E_X(\vec{p}_X) - m_N},
\end{eqnarray}
where $X$ is a hadronic state with baryon number $B = 1$ and
strangeness $S = - 1$. The lowest states contributing to the sum over
the intermediate states $X$ are $X = M B$, where $M B =
\Lambda^0\pi^0$, $\Sigma^0\pi^0$, $\Sigma^+\pi^-$, $K^-p$ and
$\bar{K}^0n$. This gives
\begin{eqnarray}\label{label2.2}
  \hspace{-0.5in}&&i\!\!\int\!\! d^4x\, \langle p
  |{\rm T}(J^{4+i5}_{50}(x)J^{4-i5}_{50}(0))|
  p\rangle =  \frac{1}{2}\sum_{\sigma_p = \pm\,1/2}\sum_{M B}
  \sum_{\sigma = \pm\,1/2}\frac{1}{32\pi^3}\nonumber\\
  \hspace{-0.5in}&&\times\,\int \frac{d^3k}{
    E_{M}(k) E_B(k)}\frac{\vert \langle M(- \vec{k} )B(\vec{k},\sigma)
    \vert J^{4-i5}_{50}(0)|
    p(\vec{0},\sigma_p)\rangle\vert^2}{E_M(k) + E_B(k)- m_N},
\end{eqnarray}
where $\vec{k}$ is the relative momentum of the $M B$ pairs. The
calculation of the matrix elements of $\langle
M(-\,\vec{k}\,)B(\vec{k},\sigma)\vert J^{4-i5}_{50}(0)|
p(\vec{0},\sigma_p)\rangle$ we carry out in the soft--meson limit (to
leading order in ChPT) using the PCAC hypothesis and Current Algebra
\cite{SA68}, and in the heavy--baryon limit, accepted in ChPT for the
analysis of baryon exchanges \cite{JG83}. This gives
\begin{eqnarray}\label{label2.3}
  \hspace{-0.3in}&&i\!\!\int\!\! d^4x\, \langle p
  |{\rm T}(J^{4+i5}_{50}(x)J^{4-i5}_{50}(0))|
  p\rangle = \nonumber\\
  \hspace{-0.3in}&&= \frac{1}{2}\sum_{\sigma_p = \pm\,1/2}
  \sum_{\sigma = \pm\,1/2}\frac{1}{32\pi^3}
  \frac{1}{F^2_M}\frac{1}{m_N}\int \frac{d^3k}{
    k^2 }\,
  \vert \langle B(\vec{k},\sigma)\vert [Q^{M}_5(0)^{\dagger},
  J^{4 - i5}_{50}(0)]\vert 
  p(\vec{0},\sigma_p)\rangle\vert^2,
\end{eqnarray}
where we have neglected the mass differences of baryons. This is valid
in the chiral limit, since the mass differences of baryons, according
to ChPT, are proportional to current quark masses and do not
contribute in the chiral limit.  $Q^{M}_5(0)^{\dagger}$ is the
axial--vector charge operator with quantum numbers of the $M$--meson
and $F_M$ is the PCAC constant of the $M$--meson: $\sqrt{2}\,F_{\pi^0}
= F_{\pi^-} = F_{K^-} = F_{\bar{K}^0} = \sqrt{2}\, F_{\pi}$.

In the framework of Gell--Mann's current algebra \cite{SA68} the
equal--time commutators $[Q^{M}_5(0)^{\dagger},J^{4 - i5}_{50}(0)]$
amount to
\begin{eqnarray}\label{label2.4}
{[Q^3_5(0), J^{4 - i5}_{50}(0)]} &=& -\,\frac{1}{2}\,J^{4 - i5}_0(0),\nonumber\\
{[Q^{1 + i2}_5(0), J^{4 - i5}_{50}(0)]} &=& J^{4 - i5}_0(0),\nonumber\\
{[Q^{4 + i5}_5(0), J^{4 - i5}_{50}(0)]} &=& J^3_0(0) + \sqrt{3}\,J^8_0(0),\nonumber\\
{[Q^{6 + i7}_5(0), J^{4 - i5}_{50}(0)]} &=& J^{1 - i2}_0(0).
\end{eqnarray}
The matrix elements of the vector currents are defined by 
\begin{eqnarray}\label{label2.5}
  \langle \Lambda^0(\vec{k},\sigma)\vert J^{4 - i5}_0(0)\vert 
  p(\vec{0},\sigma_p)\rangle &=& -\sqrt{\frac{3}{2}}\,F_V(k^2)\,
  \bar{u}_{\Lambda^0}(\vec{k},\sigma)\gamma^0 u_p(\vec{0},\sigma_p),\nonumber\\
  \langle \Sigma^0(\vec{k},\sigma)\vert J^{4 - i5}_0(0)\vert 
  p(\vec{0},\sigma_p)\rangle &=& - \sqrt{\frac{1}{2}}\,F_V(k^2)\,
  \bar{u}_{\Sigma^0}(\vec{k},\sigma)\gamma^0 u_p(\vec{0},\sigma_p),\nonumber\\
  \langle \Sigma^+(\vec{k},\sigma)\vert J^{6 - i7}_0(0)\vert 
  p(\vec{0},\sigma_p)\rangle &=& - \,F_V(k^2)\,
  \bar{u}_{\Sigma^+}(\vec{k},\sigma)\gamma^0 u_p(\vec{0},\sigma_p),\nonumber\\
  \langle p(\vec{k},\sigma)\vert J^3_0(0) + \sqrt{3}\,J^8_0(0)\vert 
  p(\vec{0},\sigma_p)\rangle &=& 2\,F_V(k^2)\,
  \bar{u}_p(\vec{k},\sigma)\gamma^0 u_p(\vec{0},\sigma_p).\nonumber\\
  \langle n(\vec{k},\sigma)\vert J^{1 - i2}_0(0)\vert 
  p(\vec{0},\sigma_p)\rangle &=& \,F_V(k^2)\,
  \bar{u}_n(\vec{k},\sigma)\gamma^0 u_p(\vec{0},\sigma_p)
\end{eqnarray}
where $F_V(k^2)$ is a form factor.  In the limit of the $SU(3)$
flavour symmetry the vector form factor $F_V(k^2)$ should be the same
for all components of the octet of ground baryons. In the
heavy--baryon limit we can drop the contribution of the ``magnetic''
form factor \cite{IV8b}. The result of the calculation of the r.h.s.
of Eq.(\ref{label2.3}) is
\begin{eqnarray}\label{label2.6}
  i\!\int\! d^4x\,\langle p |{\rm T}(J^{4+i5}_{50}(x)J^{4-i5}_{50}(0))|
  p\rangle = \frac{7 m_N }{4 \pi^2 F^2_{\pi}}\int^{\infty}_0dk\,F^2_V(k^2).
\end{eqnarray}
The contribution of the term (\ref{label2.1}) to the energy level
shift is equal to
\begin{eqnarray}\label{label2.7}
  \hspace{-0.3in} \delta \epsilon_{1s} = -\,\frac{7\alpha^3 \mu^3 m_K}{ 
    32\pi^3 F^4_{\pi}}\int^{\infty}_0dk\,F^2_V(k^2). 
\end{eqnarray}
For the numerical estimate we identify $F_V(k^2)$ with the
electromagnetic form factor of the proton $F_V(k^2) = 1/(1 +
k^2/M^2_V)^2$, where $M^2_V = 0.71\,{\rm GeV}^2$ is the squared slope
parameter \cite{MN79}. After the integration over $k$ we get
\begin{eqnarray}\label{label2.8}
  \hspace{-0.3in} \delta \epsilon_{1s} = -\,\frac{35 \alpha^3 \mu^3 m_K M_V}{ 
    1024\pi^2F^4_{\pi}} = -\,260\,{\rm  eV}.
\end{eqnarray}
Assuming that the theoretical expression for the energy level shift of
kaonic hydrogen in the ground state Eq.(\ref{label1.5}) fits the
experimental value Eq.(\ref{label1.1}) we estimate the $\sigma^{(I =
  1)}_{KN}(0)$--term. It is equal to
\begin{eqnarray}\label{label2.9}
  \sigma^{(I = 1)}_{KN}(0) = (193 + 260 - 238)\,\frac{2\pi}{\alpha^3}\,
  \frac{m_K}{\mu^3}\,F^2_{\pi} = 
  (433\pm 85)\,{\rm MeV},
\end{eqnarray}
where the uncertainty $\pm\,85\,{\rm eV}$ is defined by the
experimental and theoretical uncertainties $\pm\,37\,{\rm eV}$ and
$\pm\,21\,{\rm eV}$ of the energy level shift, respectively.
According to the proposal of the SIDDHARTA Collaboration
\cite{SIDDHARTA}, the experimental uncertainty of the energy level
shift should be substantially diminished in the new set of experiments
on the energy level displacement of kaonic hydrogen.

\section{Discussion}
\setcounter{equation}{0}

Using the theoretical value of the energy level shift of kaonic hydrogen in
the ground state, calculated within the phenomenological quantum field
theoretic model of strong low--energy $\bar{K}N$ interactions
\cite{IV3,IV8}, and the experimental data by the DEAR Collaboration
\cite{DEAR} we have extracted the value of the $\sigma^{(I =
  1)}_{KN}(0)$--term: $\sigma^{(I = 1)}_{KN}(0) = (433\pm 85)\,{\rm
  MeV}$.  The obtained result is by a factor 2 larger the estimate
$\sigma^{(I = 1)}_{KN}(0) = (200\pm 50)\,{\rm MeV}$, carried out within
ChPT with a dimensional regularization of divergent diagrams
\cite{VB93}. For the cut--off regularization of divergent integrals in
CHPT the value of the $\sigma^{(I = 1)}_{KN}(0)$--term agrees
qualitatively with our estimate (see \cite{VB93} (Borasoy)).

The value $\sigma^{(I = 1)}_{KN}(0) = (433\pm 85)\,{\rm MeV}$ agrees
well with (i) the value $\sigma_{\pi N}(0) = 61^{+2}_{-4}\,{\rm MeV}$
of the $\sigma_{\pi N}$--term of $\pi N$ scattering, extracted from
the experimental data on the energy level displacement of pionic
hydrogen in the ground state \cite{PSI2} and (ii) the vanishing
contribution of the strange quarks to the proton structure. The
$\sigma_{\pi N}$--term is defined by \cite{TE87}:
\begin{eqnarray}\label{label2.10}
  \sigma_{\pi N}(0) = \frac{m_u + m_d}{4m_p}\,
\langle p\vert \bar{u}(0)u(0) + \bar{d}(0)d(0)\vert p\rangle.
\end{eqnarray}
Following to the naive quark counting and assuming that the proton has
the quark structure $\vert p\rangle = \vert uud\rangle$ \cite{FC79},
we can calculate the matrix elements $\langle p\vert
\bar{u}(0)u(0)\vert p\rangle$ and $\langle p\vert \bar{d}(0)d(0)\vert
p\rangle$ in terms of the $\sigma_{\pi N}(0)$--term. We get
\begin{eqnarray}\label{label2.11}
  \langle p\vert \bar{u}(0)u(0)\vert p\rangle = \frac{8}{3}\,\frac{m_p}{m_u + m_d}\,
\sigma_{\pi N}(0)\;,\;
\langle p\vert \bar{d}(0)d(0)\vert p\rangle = \frac{4}{3}\,\frac{m_p}{m_u + m_d}\,
\sigma_{\pi N}(0).
\end{eqnarray}
The contribution of the strange quarks to the proton structure is
defined by the quantity \cite{JG99}
\begin{eqnarray}\label{label2.12}
y = \frac{2\langle p\vert \bar{s}(0)s(0)\vert
p\rangle}{\langle p\vert \bar{u}(0)u(0) + \bar{d}(0)d(0)\vert p\rangle }.
\end{eqnarray}
The matrix element $\langle p\vert \bar{s}(0)s(0)\vert p\rangle$ is equal to
\begin{eqnarray}\label{label2.13}
  \langle p\vert \bar{s}(0)s(0)\vert p\rangle &=& 
\frac{4m_p}{m_u + m_s}\,\Big[\sigma^{(I = 1)}_{KN}(0) -
  \,\frac{2}{3}\,\frac{m_u + m_s}{m_u + m_d}\,\sigma_{\pi N}(0)\Big] 
=\nonumber\\
&=& (-\,2.19 \pm 2.43)\times 10^3\,{\rm MeV}.
\end{eqnarray}
where we have used the current quark masses $m_u = 4\,{\rm MeV}$, $m_d
= 7\,{\rm MeV}$ and $m_s = 135\,{\rm MeV}$ \cite{JG75}, obtained at
the normalization scale $\mu = 1\,{\rm GeV}$ commensurable with the
scale of spontaneous breaking of chiral symmetry \cite{AI99}.

For the matrix elements $\langle p\vert \bar{q}(0)q(0)\vert p\rangle$,
where $q = u, d$ or $s$, defined by Eqs.(\ref{label2.11}) and
(\ref{label2.13}), we obtain $y = -\,0.21 \pm 0.23$ \cite{JG99}.  This
agrees well with the absence of the strange quarks in the proton
structure.

Since the contributions of the other states $X$ in the sum
(\ref{label2.1}) should be calculated in the chiral limit and in the
heavy--baryon limit, most of these contributions vanish. For example,
one can show that the contribution of the states $X = B M M$ vanishes
in the heavy--baryon limit as $O(1/m^2_N)$. This implies the dominant
role of the lowest states $X = BM$, which we have taken into account.

\section{Acknowledgement}

We are grateful to Torleif Ericson for helpful discussions and
constructive criticism.


\begin{thebibliography}{9}
\bibitem{IV8a} The S--wave scattering length $a^{K^-p}_0$ is defined
  by the S--wave scattering lengths $a^I_0$ of $\bar{K}N$ scattering
  with isospin $I = 0$ and $I = 1$: $a^{K^-p}_0 = (a^0_0 + a^1_0)/2$.
  In the chiral limit $a^0_0$ and $a^1_0$ satisfy the low--energy
  theorem $a^0_0 + 3 a^1_0 = 0$ with the numerical values $a^0_0 =
  (-\,1.50 \pm 0.05)\,{\rm fm}$ and $a^1_0 = (0.50 \pm 0.02)\,{\rm
    fm}$ \cite{IV8}. The contribution of inelastic channels $\delta
  a^{K^-p}_0 = ( -\,0.04\pm 0.01)\,{\rm fm}$ \cite{VI05} does not
  violate the low--energy theorem $a^0_0 + 3 a^1_0 = 0$ within
  theoretical uncertainties.
\bibitem{IV8b} The matrix element of the vector current $J^{a
    -ib}_0(0)$ between the ground baryon states can, in principle,
  contain the contribution with the structure of the magnetic moment
  \[F_M(\vec{k}^{\;2})\,\bar{u}_B\,\frac{\sigma^{0\nu}k_{\nu}}{2
    m_B}\,u_p = F_M(\vec{k}^{\;2})\, \frac{E_B - m_B}{2
    m_B}\bar{u}_B\gamma^0u_p \to F_M(\vec{k}^{\;2})
  \frac{\vec{k}^{\;2}}{4 m^2_B}\,\bar{u}_B\gamma^0u_p,\] where
    $\sigma^{0\nu} = (\gamma^0 \gamma^{\nu} - \gamma^{\nu}\gamma^0)/2$
    and $k_{\nu} = (E_B, -\,\vec{k}\,)$ and $F_M(\vec{k}^{\;2})$ is a
    ``magnetic'' form factor.  In the heavy--baryon limit $m_B \to
    \infty$ the contribution of the ``magnetic'' form factor vanishes
    as $O(1/m^2_B)$.
\end{thebibliography}
\end{document}